\documentclass[acmcsur]{acmsmall}\pdfoutput=1
\pdfoutput=1
\usepackage{etex}

\newif\ifisreport 
\isreportfalse

\usepackage{comment}

\acmVolume{online}
\acmNumber{}
\acmArticle{}
\acmYear{2017}
\acmMonth{06} 

\usepackage[T1]{fontenc}
\usepackage[utf8]{inputenc}
\usepackage{graphicx}
\usepackage{listings}
\usepackage{xcolor}
\usepackage{marginnote}

\title{Automatic Software Repair: a Bibliography}
\author{Martin Monperrus}
\date{version of March 7, 2017}

\usepackage[firstinits=true,doi=false,url=false,isbn=false,maxnames=99]{biblatex}
\DeclareFieldFormat[article,inbook,incollection,inproceedings,patent,thesis,unpublished]{titlecase}{\MakeSentenceCase*{#1}}
\DeclareFieldFormat{titlecase}{\MakeSentenceCase{#1}}
\bibliography{monperrus.bib}
\bibliography{biblio-software-repair.bib} 

\reversemarginpar 

\usepackage{pdfcomment}
\newcommand{\TODO}[1]{\textcolor{red}{todo: #1}\pdfcomment[color=yellow,open=false]{#1}}\newcommand\todo\TODO 

\usepackage{tabularx}
\usepackage{framed}
\usepackage{indentfirst}

\newcommand{\importantword}[1]{\emph{#1}\marginnote{\lowercase{#1}}\index{\lowercase{#1}}}

\ifisreport
\usepackage[hmargin=3cm,tmargin=3cm,lmargin=4cm,rmargin=3cm, marginparwidth=3.2cm, marginparsep=4mm,a4paper]{geometry} 

\long\def\tbl#1#2{
#2
\caption{#1}
}

\else
\fi

\begin{document}

\maketitle

Accepted for publication in ACM Computing Surveys on June 3rd 2017.

\medskip

\textbf{Abstract:} 
This article presents a survey on automatic software repair. Automatic software repair consists of automatically finding a solution to software bugs, without human intervention. This article considers all kinds of repair. First, it discusses behavioral repair where test-suites, contracts, models, crashing inputs are taken as oracle. Second, it discusses state repair, also known as runtime repair or runtime recovery, with techniques such as checkpoint and restart, reconfiguration, invariant restoration. The uniqueness of this article is that it spans the research communities that contribute to this body of knowledge: software engineering, dependability, operating systems, programming languages and security. It provides a novel and structured overview of the diversity of bug oracles and repair operators used in the literature.

\ifisreport
\else
\renewcommand\marginnote[1]{}
\fi

\section{Introduction}

This paper presents an annotated bibliography on automatic software repair. 
Automatic software repair consists of automatically finding a solution to software bugs\footnote{Automatic repair and tolerance against hardware bugs is out of the scope of this paper.}, without human intervention. 
This idea of automatically repairing software bugs is both important and challenging.
It is important because software has eaten the world\footnote{paraphrasing Silicon Valley's entrepreneur Marc Andreessen}, but unfortunately each bite comes with bugs. 
The software we daily use sometimes crashes, sometimes gives erroneous results, and sometimes even kills people \cite{epicfailures}. 
We do have millions of bugs in the wild, and many of them are being created every day in the new software products and releases we ship in production.
To sum up, if automatic software repair could only repair a fraction of those bugs, it would bring value to society and humanity.

 Automatic software repair is challenging because fixing bugs is a difficult task. Of course there are stupid bugs -- ``blunder'' as Knuth puts it \cite{Knuth89} -- that can be trivially fixed. However, any programmer, whether professional or hobbyist, remembers a bug that took her hours, if not days and weeks to be understood and fixed, these are the ``hairiest bugs'' \cite{eisenstadt1997my}. 
For those bugs, automatic repair is a challenging human-competitive task.

The goal of this paper is to draw the big picture of automatic software repair. In particular, it aims at presenting together the two main families of automatic repair techniques:
behavioral repair and state repair. The former is about automatically modifying the program code; the latter is about automatically modifying the execution state at runtime. 
The primary intended audience consists of researchers in computer science, with a focus on the research communities that contribute to this body of knowledge: software engineering, dependability, operating systems, programming languages and software security.
Each section also provides an introductory explanation of the key concepts behind automatic repair, which could be of high interest for practitioners and curious students.
This survey aims at covering all important works in the field or automatic software repair, with an emphasis on empiricism: the covered technique must apply to some programs done in industry and bugs that happen in practice. Works are included as follows: for each paper, the importance is qualified according to the visibility and reputation of the venue or the novelty of the idea presented in the paper. If several papers contain the same idea, only the most representative one is discussed and cited.
It is to be noted that the same concept ``repair'' has several names in the literature:
patch, fix, heal, recover, etc. Table \ref{tab:terminology} lists the main ones, as well as example notable references that use the term. In this paper, the name ``repair'' is chosen, because a program has something mechanical in nature, which fits well the daily usage of the word ``repair''. Also, it is the name used by most excellent papers in the field.

To my knowledge, there is no comparable bibliography in the literature. The look-back paper by Le Goues et al. \cite{GouesFW13} is close but only covers a fraction of the papers, and only on behavioral repair. 
On the contrary, Rinard \cite{rinard2006survival} only focused on runtime repair.
Yet, there are surveys in related fields, for instance for
fault-tolerance \cite{torres2000software},
fault localization\cite{sethi2004survey,wong2009survey},
algorithmic debugging \cite{silva2011survey} to only name a few.

To sum, the contribution of this paper is a survey on automatic software repair:
\begin{itemize}
\item This survey is across different research areas and includes contributions from the following communities: software engineering, dependability, operating systems, programming languages and software security. Similarly, it abstracts over terminology (automatic repair, self-healing, automatic recovery, etc.).
\item This survey provides the reader with an in-depth analysis of the literature according to the type of repair they perform (behavioral versus state repair) and the oracle they consider.
\end{itemize}

The remainder of this paper reads as follows.
Section \ref{sec:core-concepts} briefly presents the core concepts of automatic repair.
Section \ref{sec:behavioral-repair} discusses the main approaches of behavioral repair and
Section \ref{sec:state-repair} is about state repair.
Section \ref{sec:empirical-repair} is dedicated to the empirical works that aim at understanding the foundations of automatic repair.
Section \ref{sec:other-related-techniques} is an account on papers that are not directly about automatic repair yet have a close connection.

\begin{table}
\tbl{The diverse terminology of automatic software repair}{
\begin{tabularx}{\textwidth}{X|p{6cm}}
Expression & Example Ref. \\
\hline
automatic repair (program repair, self-repair)& \cite{LeGoues2012,logozzo2012modular,Pezze2011,Monperrus14}\\
automatic fixing (bug fixing, program fixing)& \cite{R2Fix,Wei2010}\\
automatic patching & \cite{Weimer2009,sidiroglou2005countering,lin2007autopag,Kim2013}\\
 healing (self-healing) & \cite{ghosh2007self,shaw2002self,sidiroglou2009assure} \\
automatic correction (self-correcting) & \cite{lazaar2011framework,kern2010automatic} \\
automatic recovery (self-recovering) & \cite{sozer2009flora,candea2003jagr} \\
 resilience & \cite{shaw2002self,cornu:hal-01062969}\\
 automatic workaround & \cite{Carzaniga2010}  \\
 survive (survival, survivability) & \cite{merideth2003enhancing,qin2005rx,rinard2006survival} \\
 rejuvenation & \cite{huang1995software} \\
biological metaphors: allergies, immunity, vaccination & \cite{qin2005rx,sidiroglou2005building,Locasto2006,jula2008deadlock}  \\
\end{tabularx}}
\label{tab:terminology}
\end{table}

\section{Core Concepts for Automatic Repair}
\label{sec:core-concepts}

Automatic repair is about bugs.
The literature is full of synonyms for ``bug'': defect, fault, error, failure, mistake, etc.
There are rather accepted definitions between faults, errors and failures \cite{avizienis2004basic}:
a \importantword{failure} is an observed unacceptable behavior;
an \importantword{error} is a propagating incorrect state prior to the failure (without yet having been noticed);
a \importantword{fault} is the root cause of the error (in particular incorrect code).
Although the relative clarity of those three concepts, one can hardly say that the literature, incl. the most recent papers, sticks to those definitions. 
Furthermore, if we only consider the repair literature, there is absolutely no emerging separation between ``automatic repair of failures'',  ``automatic repair of errors'' and ``automatic repair of faults''.  
However, we need a common concept for all words and in this paper, the term ``bug'' is used as an umbrella word because of its intuitiveness and wide usage, with the following definition:
\emph{A \importantword{bug} is a deviation between the expected behavior of a program execution and what it actually happened.}\footnote{Note that some authors use ``intended'' instead of ``expected'',  the latter is taken because it's really the viewpoint of the user or client that matters, not the viewpoint of the engineer who designed and developed the software.}

This definition of bug involves the notions of ``behavior'', ``execution'', ``program`` but has an implicit third subject: the observer, or reference point, that deems the behavior unexpected.
This ``observer'' can obviously  be a human user saying ``this output is not correct''. It is also classically a \importantword{specification}, in its most general meaning: a specification is a set of expected behaviors.
Specifications are polymorphic: they can be natural language documents, formal logic formulas, test suites, etc. They can even be implicit: for instance the specification ``the program shall not crash on any input'' holds for many programs while not often being explicitly written.
To some extent, the user saying ``this output is not correct'', is stating the specification on the fly. 
Consequently, automatic repair always refers to a specification and yields the following definition of automatic repair.
\emph{Automatic repair is the transformation of an unacceptable behavior of a program execution into an acceptable one according to a specification.}

A concept that is close to the one of specification is the one of \importantword{oracle}. Simply put, an oracle determines whether the result of executing a program is correct \cite{staats2011programs}.
To this extent, specification and oracle refer to the same thing: expectation, acceptability, correctness. 
However,  there is a major difference between both. An oracle is only a part of specifications, it is the part related to the expected output (when one such exists).
In addition, a specification contains information about the input ranges, about non-functional properties, etc.
For instance, a test suite is a specification, it contains test cases, which themselves contain assertions, the latter being the oracles. 

With respect to repair, the oracles can be split in two:
the \importantword{bug oracle} refers to the oracle that detects the unexpected behaviors;
the \importantword{regression oracle} refers to the oracles that check that no new bugs have been introduced during repair. 
The reason is that the program upon repair already satisfies all regression oracles, but a repair transformation may accidentally introduce a regression. 
There are more formal definitions of specification and oracle in the literature \cite{staats2011programs,survey_oracle_tse} but they do not bring much in the context of this paper.

Finally, a repair technique often targets a \importantword{bug class}\footnote{or fault class, or error class, etc.}.
A bug class is an abstract concept referring to a family of bugs that have something in common: the same symptoms, the same root cause, the same solution \cite{Monperrus2014}. For instance, well known bug classes include off-by-one errors, memory leaks, etc. However, there are many bug classes for which there are no clear definition and scope in the literature, and some of them even miss a name. While some initial taxonomies exist \cite{Tsipenyuk05,Duraes2006}, building a comprehensive taxonomy of bug classes will require years of research.

\section{Behavioral Repair}
\label{sec:behavioral-repair}

\begin{framed}
Behavioral repair consists of changing the behavior of the program under repair, i.e. changing its code. 
The modification can be done on source code, but also on binary code (e.g. Java bytecode or x86 native code).
Behavioral repair can be done offline or online at runtime. 
\end{framed}

When done offline, behavioral repair may happen in the development environment (IDE) of maintenance developers or in a continuous integration server.
Online behavioral repairs means repair done on deployed software.
Technically, behavioral repair at runtime involves a kind of dynamic software update (DSU), which is a research topic per se.

Behavioral repair involves a \importantword{repair operator}\footnote{or ``repair action''} which is a kind of modification on the program code.
For instance, one repair operator is the addition of a precondition, as shown below. 

\begin{lstlisting}
+ if (age>=18) 
    serve_adult_content()
\end{lstlisting}

The literature defines many different repair operators, that will be presented below and that are summarized in Table \ref{tab:behavioral-repair-operators}.
Sometimes the repair operator involves a \importantword{repair template}\footnote{or ``repair strategy'' or ``fix schema'' \cite{Wei2010}}, which is a parameterized snippet of code which targets the repair a specific bug class. A \importantword{repair model} \cite{Martinez2013} is a set of repair operators.

\begin{table}
\tbl{Examples of repair operators for behavioral repair}{
\begin{tabularx}{\textwidth}{X|p{6cm}}
 Operator & Example Ref.  \\
\hline
 add/remove/replace code & \cite{Weimer2009,Arcuri20113494} \\
 add a precondition  & \cite{DeMarco2014,lin2007autopag}\\
  replace a condition & \cite{DeMarco2014,nguyen13} \\
  replace assignment RHS & \cite{Gopinath,nguyen13,KaleeswaranTKO13}\\
 addition or removal of method calls & \cite{Dallmeier2009} \\
 adding a modulo for array read, truncating data for array write & \cite{lin2007autopag} \\
\end{tabularx}
}
\label{tab:behavioral-repair-operators}
\end{table}

For instance, when considering a test suite as specification \ref{sec:test-suite-repair}, a problem statement of behavioral repair is \emph{given a program and its test suite with at least one failing test case, create a patch that makes the whole test suite passing.}
This problem statement can be called test-suite based repair \cite{Monperrus2014}, and has been famously explored by Genprog, presented in Section \ref{sec:test-suite-repair}

\subsection{Repair \& Oracles}

As presented in Section \ref{sec:core-concepts}, automatic repair is with respect to an oracle.
Consequently, this section is organized according to the kind of oracle considered in the literature.

\subsubsection{Test Suites}
\label{sec:test-suite-repair}

A test suite is an input-output based specification. In modern object-oriented software, the input can be as complex as a set of interrelated objects built with a rich sequence of method calls, and the output can also be a sequence of method calls that observe the execution state and behavior in various ways.
In test-suite based repair, the failing test case acts as a bug oracle, the remaining passing test cases act as a regression oracle.

\marginnote{Genprog}Genprog is a seminal and archetypal test-suite based repair system developed at the University of Virginia ~\cite{Weimer2009,Weimer2010,forrest2009genetic}. Genprog uses three repair operators that are mutations over the abstract syntax trees (AST):
deletion of AST nodes ;
addition of AST nodes;
replacement of existing nodes.
For addition and replacement, the nodes are taken from elsewhere in the code base. This is called the redundancy assumption \cite{Martinez2014,BarrBDHS14}. Genprog is able to handle real-world large scale C code. The largest evaluation of Genprog \cite{LeGoues2012} claims that 55 out of 105 bugs can be fixed by Genprog. Those results have been later questioned, as discussed in \ref{sec:empirical-repair}. The Genprog thread of ideas yielded other papers in the original team \cite{schulte2010automated,Goues2012} and other laboratories \cite{qi2013efficientauto,qi2014strength}. 
Now that the core ideas of Genprog are well known and accepted, work needs to be done to improve the core repair operators (such as \cite{Oliveira2016}).

\marginnote{EXP}
Much before Genprog, in the mid 90ies, Stumptner and Wotawa \cite{stumptner1996model} have proposed automatic repair in a simple toy language called  EXP. The specification is a set of test cases (i.e. a test suite). To my knowledge, it is the first occurrence of test-suite based repair in the literature.

Arcuri \cite{arcuri2008automation,Arcuri2009,Arcuri20113494} defines 7 repair operators based on abstract syntax tree modification. \marginnote{Jaff} For instance, for ``promote mutation'', a node is replaced by one of its child. The operators are stacked in a random way. The prototype implementation, called Jaff, handles a subset of Java and is evaluated on toy programs.

\marginnote{Mutation} Debroy and Wong \cite{debroy2010using,nica2013use} propose to use standard mutations from the mutation testing literature to fix programs. 
Consequently, their repair models are:
replacement of an arithmetic, relational, logical, increment/decrement, or assignment operator by another operator from the same class;
decision negation in an if or while statement. Conventionally, they locate fault statements with spectrum based fault localization technique.
Nica et al. \cite{nica2013use} also use mutations for repair. Compared to Debroy and Wong, they comprehensively explore the space of all mutations.

\marginnote{Metaprogram} The key idea of Kern and Esparza \cite{kern2010automatic} is to generate a meta-program that integrates all possible mutations according to a mutation operator. The mutations that are actually executed are driven by meta-variables.  A repair is a set of values for those meta-variables. The meta-variables are valued using symbolic execution.

\marginnote{Semfix} NGuyen et al. \cite{nguyen13} proposed an approach called Semfix for repair based on symbolic execution and code synthesis. The location of the repair is found with angelic debugging \cite{chandra2011angelic}, then the repaired expression is synthesized with input-output component synthesis \cite{jha2010oracle}. The repaired locations are right hand side (RHS) of assignments and boolean conditionals, the synthesized expressions mix arithmetic and first-order logics. 
\marginnote{Angelix} One problem with Semfix is scalability. To overcome this problem, the same group has proposed Angelix \cite{mechtaev2016angelix}. Angelix is a repair system alike Sefix, where the symbolic execution phase has been seriously optimized in order to scale to large programs and obtain more than one angelic value, this is an ``angelic forest''. 

\marginnote{PAR} The PAR system \cite{Kim2013} is an approach for automatically fixing bugs of Java code.
PAR is based on repair templates: each of PAR's ten repair templates represents a common way to fix a common kind of bug. 
For instance, a common bug is the access to a null pointer, and a common fix of this bug is to add a nullness check just before the undesired access: this is template ``Null Pointer Checker''.
Some templates are parameterized by variables, for instance the ``Null Pointer Checker'' template takes a variable name as parameter.
The templates are applied and tested in a random search manner.

\marginnote{Nopol}
Nopol \cite{DeMarco2014} targets a specific fault class: conditional bugs. It repairs programs by either modifying an existing if-condition or  adding a precondition (aka. a guard) to any statement or block in the code. The modified or inserted condition is synthesized via input-output based code synthesis with SMT \cite{jha2010oracle} and predicate switching \cite{zhang2006locating}.
The Nopol system has been extended for also repairing infinite loops \cite{Lamelas2015}.

\marginnote{Relifix} Tan and Roychoudhury proposed Relifix, a repair system dedicated to fixing regression bugs \cite{relifix}.
The approach consists of 8 repair templates, some being transformation operators, the other being parameterized repair templates.
The key idea of Relifix is that the templates application are driven by the past changes, for instance, template ``add statement'' only add statements that were involved in the previous commits related to the regression.

\marginnote{DirectFix} 
Mechtaev et al. also perform test-suite based repair \cite{mechtaev2015directfix}, with the noble goal of synthesizing simple patches.
In order to do so, the assume a very specific kind of programs: those that can be expressed as  trace formulas (related to boolean programs of \cite{griesmayer2006repair}). 
Under this assumption, they can state the repair problem as a Maximum Satisfiability (MaxSAT) problem, where the smallest  patch is the one that satisfies the most constraints.

\marginnote{SPR}  SPR \cite{Long15} defines a set of staged repair operators so as to early discard  many candidate repairs that cannot pass the supplied test suite. This allows for exhaustively exploring a small and valuable search space.

\marginnote{CodePhage}
The idea of CodePhage \cite{sidiroglou15} is to transfer a check from one application to another application to avoid crashes.
The system assumes an error-triggering input that crashes one application but not the other one.
The considered errors are out of bounds access, integer overflow, and divide by zero errors.
The missing check is inferred from a symbolic expression over the input fields and validated by a regression test suite.

\marginnote{SearchRepair}
Ke and colleagues proposes SearchRepair \cite{KeASE2015}, a system inspired from code search. SearchRepair first indexes code fragments as SMT constraints, then at repair time, a fragment is retrieved by combining the desired input-output pairs and the fragments in a single constraint problem. The system is evaluated on small C programs written by students in an online course.

\marginnote{Prophet} Prophet \cite{LongProphetPOPL16} is a repair system that uses past commits to drive the repair. What is learned on past commits from version control systems is a probability distribution over a set of features of the patch. This probability distribution is then used to both speed up the repair and increase the likelihood to find correct patches. The evaluation  is done  on 69 real world defects from the Genprog benchmark, and  shows that 15 correct repairs are found. 
Le et al. \cite{Le2016HDRepair} also use history to select the most likely patch. Contrary to Prophet, the experiments were made on Java programs.

\subsubsection{Pre- and Post Conditions}

Some works use classical pre- and post-conditions \emph{à la} design-by-contract \cite{MeyerDBC} as oracle for repair.

He and Gupta \cite{He2004} use pre- and post-conditions to compute ``hypothesized program states'' (from the post condition) and ``actual program states' (from the failing input). The repair operators consist of changing the LHS or RHS of assignments, or changing a boolean condition with simple modifications (change variable, change relational operator) so that the hypothesized program state becomes compatible with the actual program state. A classical test suite is used for detecting regressions.

\marginnote{AutoFix-E} AutoFix-E is an approach by Wei et al. \cite{Wei2010,Wei2014}, it generates fixes for Eiffel programs,
relying on contacts (pre-conditions, post-conditions, invariants).
AutoFix-E uses four repair templates that consist of a snippet and an empty conditional expression to be synthesized.
The key intuition behind AutoFix-E is that both the snippet code and the conditional expression are taken from the existing contracts.

\marginnote{Alloy for repair} Gopinath et al. \cite{Gopinath} uses pre- and post-conditions written in the Alloy specification language. The function body is also translated to Alloy formulas. Then, the bounded verification mechanism of Alloy is used both to detect bugs (similar to \cite{Jackson2000}) and to identify the repair. The repair operators are changing the RHS of assignments and modifying existing if-conditions.

Könighofer and Bloem \cite{konighofer2011automated} considers assertions as specifications in programs that can be translated to SMT. The approach is static and the repair is shown to not violate the assertion for the considered input domain.
The approach is based on repair templates, such as changing the RHS of assignments or changing an arithmetic expression by a linear combination.
The templates holes are filled by the SMT solver.

\subsubsection{Abstract Behavioral Models}

An abstract behavioral model, such as a state machine encoding the object state and the corresponding allowed method calls can be used to drive the repair.

In 2006, before Genprog, Weimer \cite{weimer2006patches} proposed a first patch generation technique. It requires as input a safety policy (i.e. a typestate property or an API usage rule) and the control-flow graph of a method. The whole approach is static: the bug is detected as a static violation of the safety property, and the correctness condition of the patch is only to pass the safety check. Interestingly, the word does not mention the term ``repair'', it was not in the Zeitgeist at this time.

\marginnote{Pachika} Dallmeier et al. \cite{Dallmeier2009} presented Pachika, an approach for repairing Java programs.
The idea of Pachika is to first infer an object usage model from executions,
and then to generate a fix for failing runs in order to match the inferred expectedly correct behavior. 
The evaluation consists of fixing 18 bugs of ASPECTJ (75KLOC) and 8 of RHINO (38KLOC). 
The two repair operators of Pachika are addition and removal of method calls.
The main difference with the previous approach is that the behavioral model is mined, and not given.

\subsection{Static Analysis}

Static analysis tools outputs errors and warnings. It is possible to automatically repair them. In this case, the correctness oracle is the static analysis itself.

\marginnote{.NET}
Logozzo and Ball \cite{logozzo2012modular} proposes a repair approach on top of their static analysis toolchain for .Net code.
For a set of fault class identified statically (e.g. off-by-one errors), they propose a corresponding repair operations. 
The repair operators are specific to each fault class, for instance, it is adding a precondition, changing the size of an array allocation, etc.
The static analysis is run again to verify the correctness of the repair.

\marginnote{Arithmetic overflow} Logozzo and Martel \cite{LM13} targets a specific fault class in integer arithmetic (linear combinations). The arithmetic overflow is detected statically, and the suggested fix is a re-ordering of the arithmetic operations. The fix ensures that the overflow cannot happen anymore. On arithmetic overflows, there is also the work by Cocker et al. \cite{coker2013program}.

\marginnote{LeakFix}
Gao et al. \cite{gao2015safe}  present an approach for automatically fixing memory leaks for C programs. The approach consists of statically detecting and fixing memory leaks by inserting a deallocation statement. The evaluation is done on 14 programs in which 242 allocations are considered. 

\marginnote{DeepFix}
Gupta et al. \cite{gupta2017deepfix} devise an approach for repairing compiler errors, which is a static oracle. The originality of DeepFix is to use a language model based on deep learning to suggest fixes. They evaluate their approach by repairing student programs from an online course.

\marginnote{Buffer overflow} Muntean etl al. \cite{Muntean2015} statically detects buffer overflows. Then they have templates parameterized by a variable. The correct variable to be used in the template is found using SMT.
 
\subsection{Crashing inputs}

Behavioral repair can happen as a response to a field failure (e.g. a crashing exception or a SegFault caused by a buffer overflow). 
The repair process happens once the crashing input has been identified and minimized if possible.
The failing test case of a test suite can also be seen as a crashing input.
However, the main difference of crashing inputs and test suites from the viewpoint of oracle for repair is the following. A test suite also contains passing test cases (the regression oracle), and that failing test case contains assertions on the expected value, while crashing inputs, as their name suggests, only refer to a violation of the non-functional contract ``the program shall not crash''.

\marginnote{QACrashFix} Gao et al. \cite{QingGaoASE15} repairs crashing exceptions based on Stackoverflow. Their system, called QACrashFix, mines pairs of buggy and fixed code on Stackoverflow, in order to extracts an edit script. The edit scripts are tried in sequence in order to suppress the crashing exception.
Azim et al. \cite{Azim2014} detect field failures on Android smartphone applications. The considered faults are unhandled exceptions, the repair operator consists of adding try/catch blocks with binary rewriting.
\marginnote{Clotho}
Clotho \cite{dhar2015clotho} is a system that generates simple catch blocks to handle certain runtime exceptions related to string manipulation in Java. The content of the catch block is based on constraints that are collected both statically and dynamically.

\marginnote{Overflows} Sidoroglou and Keromytis \cite{sidiroglou2005countering} detect buffer overflow vulnerabilities at runtime in production, then they obtain the source of the vulnerability through the use of ProPolice \cite{eto2001propolice}; finally, they use code transformation rules written in the transformation language TXL to modify source code. Regressions are caught by manually provided  test suites.

\marginnote{AutoPag} Lin et al. \cite{lin2007autopag} tries to generate a source code patch from a working exploit that triggers an array overflow in C code. Its repair operators consist of fixing out-of-bound reads by adding a modulo in the read expression and out-of-bound writes by truncating data to be written (similarly to failure-oblivious computing).

\marginnote{SoupInt} Wang et al. \cite{wang2014diagnosis} target automatic repair of integer overflows. They have three repair operators. The first one is to force taking an error branch before the overflow happens, the second one is to  force taking an error branch after the overflow has happened, and the last one is a program stop (exit). The generated conditions are path conditions obtained from dynamic symbolic execution.

\subsection{Other Oracles}

Other specific oracles have been used in an automatic repair setting.

A number of techniques have been proposed to fix concurrency bugs.
\marginnote{AFix} Jin et al. \cite{jin2011automated} present AFix: the repair model of AFix consists of putting instructions into critical regions. This work on automatic repair of concurrency bugs has been further extended \cite{Liu2012}. Lin et al. \cite{lin2014automatic} also insert locks by encoding the problem as a satisfiability one.
\marginnote{DFixer}In Dfixer \cite{cai2016fixing}, no new locks are introduced to repair concurrency bugs, instead existing locks are pre-acquired in one thread.
\marginnote{HFix}More recently, Liu et al. \cite{liu2016understanding} have proposed another repair operator for concurrency bugs in a tool called HFix: they propose to automatically add thread-join operations.

\marginnote{malformed HTML} Samimi et al. \cite{SamirniSAMTH12} have presented an approach for repairing web application in PHP that generates HTML tags. 
The oracle that is used is whether the output HTML string is malformed, i.e. that it does not contain a inconsistent sequence of opening and closing tags (e.g. ``\texttt{<a></i></a>}''). They encode the repair as a constraint problem on strings. Wang et al.  \cite{wang2012presentationchanges} also repairs the HTML code output by PHP code, using runtime tracing instead of constraint solving.
Medeiros et al. \cite{medeiros2014repairweb} also repairs web applications, but consider SQL injection, and their repair operator consists of wrapping certain call by a sanitization function.

\marginnote{R2Fix} Liu et al. \cite{R2Fix} uses as oracle a manually written bug report. The have parameterized repair templates and extract the actual value of the template parameter from the bug report. For instance, for a not-null checker template, they extract the name of the variable to be checked from the bug report. 

\marginnote{Proof}Dennis et al. \cite{dennis2006proof} uses proof-based program verification on ML programs using Isabel as oracle. When the proof fails, the counter-example of the proof drives a repair approach based on repair templates (replacing one method call by another, adding some code). 

\marginnote{Reference implementation}
It is possible to use a reference implementation as specification for repair. 
In this case, the reference implementation both acts as the bug oracle (when the behavior of the reference implementation and of the buggy program do not correspond) and as a regression oracle. This has been little explored in the context of repair.
The approach by Könighofer and Bloem \cite{konighofer2013repair} uses SMT-based templates.
The approach by Singh et al. \cite{singh2013automated} is conceptually similar but is realized differently and the evaluation is much larger. The reference implementation and the program to be repaired are written in Python. The system translates them to a programming environment called Sketch, which is responsible for exploring the space of candidate fixes. The evaluation is made on thousands of buggy programs submitted for an online course.
Qlose \cite{dantoni2016qlose} is a similar approach based on Sketch, the novelty of Qlose is that it tries to semantic impact of the repair, by minimizing the number of inputs for which there is a behavioral change.

\marginnote{Metamorphic relations}
Jiang et al. \cite{jiang2016metamorphic} have proposed to use metamorphic relations as repair oracle. They evaluate their approach on the Introclass benchmark made of student programs. Due to the limited size of their experimental subjects, it is yet to be proven that metamorphic relations can help repair large and real programs.
Kneuss et al. \cite{kneuss2015deductive} use a kind of symbolic tests for repairing a purely functional toy language. As metamorphic relations, the symbolic tests enable to generate new test data.

\subsection{Domain Specific Repair}

The concept of automatic repair can be applied on many computational artifacts.
Indeed, there are many works doing automatic repair in contexts that are specific to an application domain.

Lazaar et al. \cite{lazaar2011framework} repair constraint programs. With a domains-specific fault localization strategy, the repair consists of removing or adding new constraints.
Gopinath et al. \cite{gopinath2014data} repair database selection statements in a specific data-oriented language called Abap. 
Kalyanpur et al. \cite{Kalyanpur2006} state an automatic repair problem in the context of OWL ontologies.
Griesmayer et al. \cite{griesmayer2006repair} repair a specific class of programs called boolean programs: those that only contain boolean variables.
Further work has been done on repairing boolean programs \cite{samanta2014cost}.
Son et al. \cite{son2013fix} repairs access-control policies in web applications, using a static analysis and transformations tailored to this domain.

\marginnote{architectural repair}
Nentwich et al. \cite{Nentwich2003} detect inconsistencies and propose repair actions on XML documents. Their approach is applicable to all structured documents with explicit static inconsistency rules.
Along the same line, Xiong et al. \cite{xiong2009ModelInconsistencyFixing} detect and fix inconsistencies in MOF and UML models
; da Silva \cite{Silva2010} use Prolog to propose a repair plan that fixes inconsistencies in UML models;
Xiong et al. \cite{xiong2015rangefixes} focuses on automatically repairing configuration errors in software product lines.

Tran et al. \cite{Tran2000} uses repair in the sense of forcing a match between source code dependencies and a dependency model that specifies the acceptable dependencies; this can be called ``architectural repair''

\marginnote{test repair}
The approach of Daniel et al. \cite{ReAssert09} does not repair programs but the test cases that are broken in the presence of refactoring.
Memon \cite{memon2008automatically} and Gao et al. \cite{Gao2015} repair GUI test scripts. For instance, the approaches change the identifiers that are used for driving the GUI manipulation.
Leotta et al. \cite{leotta2013repairingselenium} do test repair in the context of Selenium tests, which are tests for web applications with HTML output.

\subsection{Fault Classes and Repair}
\label{sec:fault-class-repair}

Some fault classes are well-enough understood so that one can write a code transformation that suppresses all instances of the fault class at once.
For instance, one can transform 64-bit integers to unlimited precision arithmetic objects (such as BigInteger in Java) to avoid all arithmetic overflows.
In  the related work, most repair transformations for fault classes are semantic-preserving, but not necessarily.

\marginnote{failure-oblivious computing}
For instance, a seminal work on semantic modifying transformations is failure-oblivious computing \cite{rinard2004enhancing}.
Considering erroneous reads out of the bounds of an array, failure-oblivious computing transforms the code so that the read returns either the first non-null element, or the element modulo the length of the allocated array. 
Along the same line, Rinard et al. \cite{rinard2004dynamic} proposes that out-of-bounds writes are stored in a hashtable and that subsequent reads to the out-of-bound index return the object previously stored in the hashtable.
This line of research is based on the philosophical foundation than acceptable results is more important than correct results, this is called ``acceptability-oriented computing'' \cite{rinard2003acceptability}.

\marginnote{SQL injection} 
Thomas and Williams \cite{thomas2007using} propose an approach to automatically transform PHP code to secure SQL statements. The transformations modify the abstract syntax trees in order to inject secured ``prepared statements''.

\marginnote{Findbugs} 
At Google, they develop and use a tool called ``error-prone' '\cite{Aftandilian2012}, it does automatic repair of Findbugs like errors \cite{Findbugs:04:HovemeyerPugh}.
Lawall et al. \cite{lawall2009wysiwib} also defined an approach for declaratively specifying bug patterns and the corresponding patches in a tool called Coccinelle.
The same idea has been developed by Kalval and Warburton \cite{kalvala2011formal} where the repair strategy is written using a formal transformation language called Trans.

Shaw et al. \cite{shaw2014automatically} describe two transformations to fix C buffer overflows: replacement of unsafe calls by alternative safe libraries and replacement of unsafe types by safer ones.
They show that the transformations scale to large programs, do not break the existing tests and do not slow down the programs. 
Coker and Hafiz employ a similar approach for another fault class: integer arithmetic bugs \cite{coker2013program}. They propose three program transformations dedicated to integers, and show that the approach scales to real programs.

\marginnote{input filter generation}
Long et al. \cite{long2014sound} uses a static analysis specific to integer arithmetic that detects integer overflow.
For all detected potential overflows, the system infers a filter that simply discards the input. To this extent, the repair action is denying the input, a technique also done at runtime and discussed in Section \ref{sec:input-modification}.

\marginnote{unhandled exceptions}
Cornu et al. \cite{cornu:hal-01062969} target unhandled exceptions in Java. They analyze test suite executions to identify the good catch blocks that have resilience capabilities. Then, they transform the caught exception type into a more generic one (i.e. a superclass exception) so as to catch exceptions that would not be caught otherwise. The code transformation, called ``catch stretching'' is a kind of proactive repair against unexpected exceptions.

\section{State Repair}
\label{sec:state-repair}

\begin{framed}
State repair consists in changing the state of the program under repair. The state is meant in its largest acceptation: it can be changing the input, the heap, the stack, the environment. 
For instance, automatic breaking a cycle in a linked list is one kind of state repair.
As opposed to behavioral repair, state repair is necessarily done at runtime. 
\end{framed}

State repair can be rooted in classical fault tolerance \cite{avizienis2004basic}. 
In this large research field, much research has targeted ``recovery'', which  Avizienis et al. defines as 
transforming \emph{``a system state that contains one or more errors and (possibly) faults into a state without detected errors''} \cite{avizienis2004basic}.
In this paper, the term``state repair'' is used instead of ``recovery''.
This terminological move allows to have an umbrella term, ``repair'' above intrinsically related concepts (recovery, resilience, etc), and above behavioral and state repair, see Figure 5.1 of \cite{avizienis2004basic} for a bird's eye presentation of classical recovery, error handling and fault handling.

State repair requires an oracle of the bug, an oracle of incorrectness.
As opposed to behavioral repair, those oracles have to be available in production, at runtime.
This rules out certain oracles discussed in Section \ref{sec:behavioral-repair}, such as test suites, and oracles based on static analysis.
For state repair, there are three main families of bug oracles.
\marginnote{contract}
First, state repair often considers violations of non-functional contracts.
For instance, crashing with a Segfault or a null pointer exception violates the non-functional contract ``the program shall never crash''.
Second, state repair can also consider functional contracts that are verifiable in production such as pre- and post-conditions. This will be much discussed in Section \ref{sec:invariant-restoration}.
Third, there are state repair approaches that reason on ``inferred contracts'', obtained by observing the regularities of program states at runtime. In this case,  a bug is defined as a program state or behavior that violates those inferred contracts and repair is a follow-up of anomaly detection on program states and executions.

In the following,  the approaches are ordered by repair operators. This more fits to the history of the field than the ordering by kind of oracles, as what was done for behavioral repair.

\subsection{Reinitialization \& Restart}

Restarting (aka rebooting) a software application is the simplest repair action.
\marginnote{Software Rejuvenation}
It has been much explored under the term ``software rejuvenation'' \cite{huang1995software}, but rather with a theoretical stance rather than a practical one.

\marginnote{Microreboot}
Candea and colleagues \cite{candea2001recursive,candea2003jagr,Candea2004} explored in depth the concept of microreboot. Microreboot consists of having a hierarchical structure of fine-grain rebootable components, and, in the presence of failures, to try to restart the application from the smallest component (an EJB) to the biggest one (the physical machine) (in a way that is similar to progressive retry in distributed computing \cite{wang1997progressive}). Their experiments show that this can significantly improve the availability of systems.

\subsection{Checkpoint \& Rollback}

\marginnote{Checkpoint}
A checkpoint and rollback mechanism takes regular snapshots of the execution state and is capable of restoring them later on.
The challenges of checkpoint and rollback are first the size and boundaries of the captured state and second the point in time of checkpointing \cite{koo1987checkpointing,kasbekar1999selective}.
When a system is equipped with a checkpoint and rollback mechanism, the rollback is the repair.
Despite being an old technique, it is valuable in a number of contexts.

Dira \cite{smirnov2005dira} is a system that instruments code to detect and recover from control-hijacking attacks through malicious payloads. The repair consists of finding the least common ancestor of the function in which the attack is detected and the one in which the payload was read in. Then, the execution is resumed to this frame and all state changes are undone.  
Similarly, Assure \cite{sidiroglou2009assure} is a technique also based on checkpointing to provide self-healing capabilities. 
Recent papers also do checkpoint and rollback as part of the repair, such as  \cite{Carzaniga2013}.

\begin{table}
\tbl{Examples of state change operators for runtime repair}{
\begin{tabularx}{\textwidth}{X | p{6cm}}
 Operator & Example Ref.\\
\hline
 restart & \cite{candea2001recursive,wang1997progressive}\\
 try an alternative implementation & \cite{avizienis85,randell75,Carzaniga2013} \\
 modify the input & \cite{ammann88,Long2012,liang2005fast} \\
 simulate a known error (aka error virtualization) & \cite{sidiroglou2009assure,Locasto2006} \\
 change the execution environment & \cite{qin2005rx,nguyen2007detecting,garvin2011using}\\
\end{tabularx}
}
\label{tab:state-repair-operators}
\end{table}

\subsection{Alternatives}

\marginnote{recovery block}
Another classical concept of fault-tolerance is n-version programming. Either with voting \cite{avizienis85} or retrying with recovery blocks \cite{randell75}, it consists of relying on alternative implementations to recover from errors.
This concept is now explored using natural sets of alternatives (as opposed to being engineered) or with automatically created sets of variants \cite{evol-design-patching2013}.
\marginnote{automatic workaround}
For instance, Carzaniga et al. \cite{Carzaniga2010} repair web applications at runtime with a repair strategy that is based on a set of API-specific alternative rules: for instance calling \texttt{bar()} instead of \texttt{foo()}. They later applied the same idea for recovering from runtime exceptions in Java \cite{Carzaniga2013}.
Hosek and Cadar \cite{hosek2013safe} use a different kind of natural diversity: upon failures, they switch from past or newer versions of the same application. The key idea is that bugginess is not monotonic: some bugs disappear while others appear over time.

\subsection{Reconfiguration}
\label{sec:repair-reconfiguration}
Reconfiguring an application is one kind of recovery \cite{avizienis2004basic}, thus one kind of state repair.
Indeed, it has much been explored when ``self-healing'' was a hype term.
For instance, Cheng et al. \cite{Cheng2002} use the three core runtime reconfiguration operators (add component, move component, delete component) to optimize quality-of-service values.
The same line of repair can be found in \cite{garlan2003increasing,sicard2008using}, which are relatively cited papers.
In the context of web service orchestration, the repair actions of Friedrich et al. \cite{Pernici2007,Friedrich2010} consist of substituting it a web service by another (which is a reconfiguration) and retrying a service call.

\subsection{Input Modification}
\label{sec:input-modification}

If the system fails on some input, one state repair action consists of modifying the input. Denying the input is also a possible option, which can be considered as an extreme case of input modification.

\marginnote{data diversity}
Ammann and Knight's ``data diversity'' \cite{ammann88} aims at enabling the computation of a program in the presence of failures. The idea of data diversity is that, when a failure occurs, the input data is changed so that the new input resulting from the change does not result in a failure. The assumption is that the output based on this artificial input, through an inverse transformation, remains acceptable in the domain under consideration.

\marginnote{input rectification} Long et al \cite{Long2012} present the idea of automated input rectification: instead of refusing anomalous inputs, they change it so that it fits into the space of typical and acceptable inputs, this is called ``input rectification''.

Liand and Sekar \cite{liang2005fast} repair buffer overflows by learning common profiles between the characteristics of crashing inputs. 
Once a valid profile is identified, crashing inputs are denied.
While the paper is about security, it can be seen as a runtime technique to repair memory errors of the form of buffer overflows. The fact that the buffer overflow is accidental (due to a bug) or maliciously triggered is irrelevant from a repair perspective.
Along the same line of input denying, Vigilante \cite{costa2005vigilante} is an integrated approach for mitigating malicious attacks. The counter-measure to worm attacks is filtering: once invalid or malicious inputs are detected they are filtered out and the current request or task is aborted.

\subsection{Environment Perturbation}

If the system fails under certain conditions, one can get the next requests to succeed by changing the runtime environment (e.g. the memory, the scheduling) or the configuration.

\marginnote{Rx}Qin et al. \cite{qin2005rx} shows that memory errors can be avoided by padding allocated memory blocks with extra space. 
\marginnote{DieHard} Berger and Zorn \cite{berger2006diehard} do the same thing and add replication. However, the difference with Rx is that their system allows for probabilistic reasoning on the resulting memory safety.
\marginnote{Exterminator}Novark et al. \cite{novark2007exterminator} explores the same idea.
Differently, Nguyen and Rinard \cite{nguyen2007detecting} enforces a bounded memory size by cyclic memory allocation  in a way that is similar to failure-oblivious computing (already presented in Section \ref{sec:fault-class-repair}). 
Garvin et al. \cite{garvin2011using} address configuration bugs and propose ``reconfiguration workarounds'' that change the configuration causing a failure.

Jula et al. \cite{jula2008deadlock} presents a system to defend against deadlocks at runtime. The system first detects synchronization patterns of deadlocks, and when the pattern is detected, the system avoids re-occurrences of the deadlock with additional locks. 

Tallam et al. \cite{Tallam2008avoiding} names this family of technique ``execution perturbations''. 
For concurrency and memory bugs, they show that removing thread interruptions, padding memory allocations, and performing denial of requests is a way to avoid failures.

\subsection{Rollforward}

Rollforward (or forward recovery) means transforming the current system state into a correct one.
There are several techniques of forward recovery: invariant restoration, error virtualization, etc.

\subsubsection{Invariant Restoration}
\label{sec:invariant-restoration}
In some cases, state correctness can be expressed as an invariant. 
Consequently, repair means restoring the invariant, if possible with a minimum of changes from the current erroneous state.

Demsky and Rinard \cite{demsky2003automatic} uses a specification language to express correctness properties on data structures. This specification is then used at runtime to automatically repair broken data structure (concrete instances at runtime, not the abstract data type).
Elkarablieh et al. \cite{elkarablieh2007assertion} also automatically repair data structures at runtime, the difference with Demsky and Rinard is that they rely on an invariant written in regular Java code (a ``repOK'' boolean method).

\marginnote{ClearView}Perkins et al. \cite{Perkins2009} presented ClearView a system for automatically repairing errors in production. The system works on low level x86 binaries and consists of monitoring the system execution to learn invariants. Those invariants are then monitored, and a violation is followed by a forced restoration. The repairs are at the level of CPU registers and memory location changes.

Lewis and Whitehead~\cite{Lewis2010} have a generic repair approach for event-based system by defining a runtime fault-monitor, but the core idea is that same: when an invariant is violated, the repair system automatically restores it. The example in a video-game domain is fun:
if Mario is hanged in the sky due to specific sequence of actions and interactions, it is forcefully put back on the ground. Beyond data structures and video-games, in real systems, many strange and undesired system states can happen from complex chains of events and interactions, but it is often possible to state simple invariants to guide runtime repair.

\subsubsection{Error virtualization}

Error virtualization consists of handling an unknown and unrecoverable error with error-handling code that is already present in the system yet designed for handling other errors.

This idea has been much explored at Columbia University.
For instance, Sidiroglou et al. \cite{sidiroglou2005building} do error virtualization in system that imitates biological immunity. They combine error virtualization with selective transactional emulation,  a technique consisting of emulating the execution of native code with an interpreter in a transactional manner.
When a failure occurs in an emulated section, all state changes are undone (a kind of micro rollback at the level of functions).
\marginnote{Assure}In Assure \cite{sidiroglou2009assure}, the idea of error virtualization is associated with fuzzing to discover and test in advance valuable error virtualization points, called rescue points.

\subsubsection{Other Forward Recovery}

Carbin et al. \cite{Carbin2011} introduced a system that monitors programs in order to detect infinite loops and escaping them. The system works with binary code instrumentation and breaks the loops with no memory state changes detected during their execution.
\marginnote{loop perforation}Along the same line is the concept of ``loop perforation'' \cite{Sidiroglou-Douskos2011}.
Sidiroglou et al. have shown \cite{Sidiroglou-Douskos2011} that it is possible to skip the execution of loop iterations in certain application domains. For instance, in a video decoding algorithm (codec), skipping some loop iterations only has an effect on some pixels or contours but does not completely degrade or crash the software application. On the other hand, skipping loop iterations is key with respect to performance. In other words, there is a trade-off between the performance and accuracy. This trade-off can be set offline (e.g. by arbitrarily skipping one every two loops) or dynamically based on the current load of the machine.

Dobilyi and Weimer \cite{dobolyi2008changing} target repair of null pointer exceptions. Using code transformation, they introduce hooks to a recovery framework. This framework is responsible for forward recovery of the form of creating a default object of an appropriate type to replace the null value or of skipping instructions.

\marginnote{RCV} Long et al. \cite{LongSR14} introduces the idea of ``recovery shepherding''.
Upon certain errors (null dereferences and divide by zero), recovery shepherding consists in returning a manufactured value, as for failure oblivious computing. However, the key idea of recovery shepherding is to track the manufactured value so as to see 1) whether they are passed to system calls or files and 2) whether they disappear. In the former case, system calls and file writes are disabled if they involve a fake manufactured value, in order to limit error propagation. When a manufactured value is no longer used and referenced, it means that the error has somehow evaporated, and the experiments of the paper show that this is often the case.

\subsection{Collaborative Repair}

A cross-cutting concern of repair at runtime is to share the repairs that work across all instances of the same application.
This has been explored under the name of ``application community''.
Locasto et al. \cite{Locasto2006} uses application communities to find and distribute repairs of the form of stack manipulation. 
Rinard et al. \cite{rinard2011collaborative} also reports on experiments on the centralization of monitoring information and the distribution of repairs across a community of applications.

\section{Empirical Knowledge on Repair}
\label{sec:empirical-repair}

Beyond proposing new repair techniques, there is a thread of research on empirically investigating the foundations, impact and applicability of automatic repair, whether behavioral or state repair. 

There is wealth of information in software repositories that can be used for repair.
In particular, one can mine bug reports and commits for knowledge that is valuable for automatic repair.
Martinez \& Monperrus \cite{Martinez2013} studied $89,993$ of commits to mine repair actions from manually-written patches. By repair actions, they mean kinds of changes on the abstract syntax trees of programs such as modifying an if condition.
They later investigated \cite{Martinez2014} the \importantword{redundancy assumption} in automatic repair (whether you can fix bugs by rearranging existing code), and found that it holds in practice: many bug fix commits only rearrange existing code, a result confirmed by Barr et al. \cite{BarrBDHS14}. 
Zhong \& Su \cite{zhong2015an} conducted a case study on over 9,000 real-world patches and found important facts for automatic repair: for instance, their analysis outlines that some bugs are repaired with changing the configuration files.

On the goodness of synthesized patches,  Fry et al. \cite{FryLW12humanstudy} conducted a study of machine-generated patches based on 150 participants and 32 real-world defects. Their work shows that machine-generated patches are slightly less maintainable than human-written ones. Tao et al. \cite{TaoKKX14} performed a similar study to study whether machine-generated patches assist human debugging. Monperrus \cite{Monperrus2014} further discussed the \importantword{patch acceptability} criteria of synthesized patches and emphasized that assessing patch acceptability may require a high level of expertise, a result confirmed by \cite{martinez2016}. Qi et al. \cite{qi2015kali} are the first to thoroughly analyze the patches generated by Genprog, and found that most of them are incorrect. It is an open question whether this holds for test-suite based repair in general or not \cite{martinez2016}.
When they are incorrect, it is because they exploit specificities and weaknesses of the test suite, which can be seen as a kind of \importantword{overfitting}.
A repair technique is said to overfit when the synthesized patch only works on the failing inputs and fails to generalize. 
Smith et al. \cite{Smith15fse} also studied the problem of overfitting in automatic repair; on a dataset of student programs, they show that Genprog and related techniques do suffer from overfitting.

A study by Kong et al. \cite{kong2015experience} compares different repair systems: GenProg \cite{LeGoues2012}, RSRepair \cite{qi2014strength}, and AE \cite{weimer2013leveraging}. They report repair results on 119 seeded bugs and 34 real bugs from the Siemens benchmark, and show that not all techniques are equal.

\marginnote{benchmark}
Finally, for the knowledge on repair to consolidate, there is a need for accepted, well-defined and publicly available benchmarks \cite{Monperrus2014}. Le~Goues et al. \cite{LeGoues15tse} have set up such a benchmarks for bugs in C programs, it totals 1183 bugs, collected in open-source projects and student code.

\section{Related Techniques}
\label{sec:other-related-techniques}

We now present works that are related to automatic repair, yet not being ``automatic repair'' per se, according to the definitions we gave in Section \ref{sec:behavioral-repair} and \ref{sec:state-repair}.
In particular, they either miss the full automation or the actual repair of real programs.

\subsection{Forward Engineering For Repair}

Many authors have tried to list the important principles to have robust, resilient if not self-repairable applications.
These principles can be implemented and enforced as first-class concepts in frameworks and libraries.
This is what can be called ``forward engineering  for repair''.

Somayaji et al. describe principles to build immune computer systems \cite{somayaji1998principles}:
distributability, multi-layering, diversity, disposability, autonomy, adaptability, behavioral sense-of-self, anomaly detection.
Candea and Fox \cite{candea2003crash} define a set of characteristics for programs to recover quickly: with those characteristics an application becomes ``\importantword{crash-only software}''. The two key characteristics are that all interactions between components have a timeout and all resources are leased.
Sussmann \cite{sussman2007building} as well as Gabriel and Goldmann \cite{gabriel2006} also provide insightful perspectives on how to build resilient and self-repairable software.

There are also frameworks for supporting repair.
Flora \cite{sozer2009flora} is a framework to support local restart in applications. It is principally composed of a communication manager for dropping or queuing messages between components.
Denaro et al. \cite{Denaro2009} proposes an architecture to fix interoperability bugs in service oriented systems.
Adaptors between service variants are manually written and are selected at runtime to enable correct communication.
Levinson \cite{levinson2005unified} defines an embedded DSL to support runtime searches in a space of program variations.
\marginnote{SafeDrive}Zhou et al. \cite{zhou2006safedrive} defines annotations for operating system C code in order to recover from driver errors in Linux. The annotations are checked by a type system and drives invariant restoration.
\marginnote{Bristlecone}Demsky and Dash \cite{demsky2008bristlecone} proposes Bristlecone, a language with built-in robustness capabilities. Bristlecone is based on tasks and dependences between tasks, as well as transactional state changes. Error-handling is thus fully automated.

A known characteristic of bugs is that the same kind of bug can affect many different locations in the same code base.
In this case, it is desirable to write a unique patch that is then applied to all those locations.
The generic patch can be inferred from a concrete instance at a given location or written in an abstract way.
\marginnote{systematic edit}
This has been called ``systematic editing'' by Meng et al. \cite{Meng2011}.
Similarly, Sun et al. \cite{sun2008automated,sun2010propagating} propose tool support for patch applications. 
\marginnote{Coccinelle}
The Coccinelle tool \cite{padioleau2008documenting}  also provides this functionality. 
The abstract patches can be automatically inferred from concrete instances \cite{andersen2010generic,Meng2013}.

\subsection{Repair Suggestions}
\label{sec:repair-suggestions}

There are some systems which give ``repair suggestions'' to the developer. 
While it is not fully automated, if the suggestion is correct, such a system can be seen as providing partial automatic repair, where the repair system and the developer work in tandem.

\marginnote{HelpMeout}
Hartmann et al. \cite{hartmann2010would} designed  a system called ``HelpMeout'' that proposes suggestions to fix error messages. The system targets compiler error messages and runtime exceptions. It first collects error messages and the associated changes that occur on developer's machines that are monitored. Then, when the same error message is encountered by another developer, the system compares the erroneous source file with the closest fixed version that is in the database. It uses a tailored distance metric to increase the relevance of suggestions.

\marginnote{BugFix} Jeffrey et al. \cite{Jeffrey2009} presented a fix suggestion approach based on association rules. The rules suggest a bug fix action for suspicious statements represented by a number of features (in the machine learning meaning). 
The features (called ``descriptors'' in the paper) are abstraction over the tokens of the statements. The prediction also uses ``interesting value mapping pairs'' (IVMP) which are concrete values that enable test cases to pass (aka value replacement \cite{jeffrey2008fault} and angelic values \cite{chandra2011angelic,DeMarco2014}).
The bug fix recommendations are typical comparison operator change, constant change, add or increase numerical values. 

\marginnote{MintHint} Kaleeswaran et al. \cite{KaleeswaranTKO13} have proposed a repair suggestion approach based on correlations variable values and expected output. The expected output is obtained through concolic executions, and the repair hints consist of changing the RHS of a single assignment statement.

\marginnote{Excel}Abraham and Erwig \cite{Abraham2005} suggest change in Excel formulas.   
Malik et al. \cite{Malik2011} transform runtime data structure repair (see \ref{sec:invariant-restoration}) as fix suggestions.
Brodie et al. \cite{brodie2005quickly} design a distance metric across call stacks (stack trace) to match issue reports and known fixes.

\subsection{Theoretical Software Repair}

Some authors explore automatic repair with strong assumptions under which there exists no program in practice.
To the best of our knowledge, there is no survey paper on this area, but the article by Bodik and Jobstmann contains a dedicated section about this
\cite{Bodik2013}. Here, the most notable papers in this area are briefly mentioned for giving the reader a first set of pointers.
For instance, Jobstmann et al. \cite{Jobstmann2005} repair programs that are expressed in linear temporal logics.
George \cite{George2003biological} describes a simple and theoretical programming model that supports automatic recovery via a kind of  homeostasis that maintains invariants.
Fisher et al. \cite{fisher_broken_program} also perform repair on a toy formal language.
Wang and Cheng \cite{wang2008suggestion} state program repair as edit sequences on state machines.
Zhang and Ding  \cite{zhang2008ctl} repair  computation tree logic models.

\section{Conclusion}

This article has presented an annotated bibliography on automatic software repair. 
This research field is both old and new.
It is old because we can find techniques related to automatic repair in fault-tolerance papers from the 70es and 80es, for instance a 1973 paper
is entitled ``STAREX self-repair routines: software recovery in the JPL-STAR computer'' \cite{starex1973}.
It is new, because the idea of automatically changing the code, i.e. behavioral repair, has started to be explored only since the end of 2000.
Whether old or new, the techniques have to scale to today's size and complexity or software stacks, and we are not there yet.
This means that this is only the beginning, and in the upcoming years, we are going to have much fun, surprise and admiration in the field of automatic software repair.

\ifisreport
\else
\printbibliography
\fi

\end{document}